\documentstyle[12pt,sprocl]{article}

\newcommand{\be}{\begin{equation}}
\newcommand{\ee}{\end{equation}}
\newcommand{\bear}{\begin{eqnarray}}
\newcommand{\eear}{\end{eqnarray}}

\begin{document}

\vspace{0in}
\begin{flushright}
FERMILAB-CONF-96/019-T
\end{flushright}
\vspace{.7in}

\centerline{\normalsize\bf LONGITUDINAL Z-BOSON PAIR PRODUCTION VIA}
\baselineskip=22pt
\centerline{\normalsize\bf GLUON
FUSION IN TECHNICOLOR MODELS \footnotesize\sf\footnote{Talk presented at International symposium on particle
theory and phenomenology, ISU, Ames, 1995}
}
\baselineskip=16pt
\centerline{\normalsize\bf }
\centerline{
 TAEKOON LEE}
\baselineskip=13pt
\centerline{\footnotesize\it Fermilab, P.O.Box 500, Batavia, IL 60510}

\vspace*{0.9cm}
\abstracts{We study the coupling of two longitudinal Z-bosons to two gluons via
technicolor interactions. 
Noticing a similarity of this process to
$2 \gamma \rightarrow 2 \pi^{0}$, we calculate the amplitude in chiral 
perturbation theory in one-generation technicolor model. At
the invariant mass of $
Z_{L}Z_{L}$ 
above the colored pseudo-Goldstone boson threshold, we find the signal
in proton collisions at $\sqrt{s} = 14\, TeV$
is stronger than the standard model background by an order of magnitude,
and is large enough to be easily observable at LHC.
  The cross section for $W_{L}^{+}W_{L}^{-}$ pair
production is also presented.}

\normalsize\baselineskip=15pt
\setcounter{footnote}{0}
\renewcommand{\thefootnote}{\alph{footnote}}
\section{Introduction}
The standard model of electroweak interactions is remarkably good
in explaining numerous electroweak phenomena. However, one of
its most important part, the symmetry breaking sector(SBS), remains a
mystery, and its full understanding is essential to complete the theory.
Numerous models were proposed as a candidate for SBS. Notable 
among them are the well-known minimal Higgs model, SUSY motivated
models, and dynamical models such as  Technicolor. 
It is therefore important to find effective probes of SBS and 
identify the right symmetry breaking mechanism. Among the
many ideas proposed for this purpose, well-known are the precision
measurement of the standard model, and the direct scattering of
longitudinal weak gauge bosons. From the former, simplest TC models
with large number of technicolors or families are already 
excluded. 

The purpose of this talk is to show that the process $ gg 
\rightarrow Z_{L}Z_{L}$
in proton collisions could be an effective probe of TC-type SBS.
 The main advantage of this process 
is that  due to its large cross section it can be
easily observable  at LHC with very clean signal such as four leptons of
electrons and muons.
In a generic one-generation TC model, we find the signal is stronger
than the standard model background by an order of magnitude. 
This process with an arbitrary Z-boson polarization  
 was investigated in the standard model  by Glover and Van Der Bij \cite{R1}. It
was shown that a Higgs resonance with not too heavy mass could
be easily observed  through this process at proton collisions at
LHC.  Combining our result with Glover and Van Der Bij's, we think the
Z-boson pair production via gluon fusion can be an effective
probe of SBS, either elementary Higgs or TC type.

We begin our study by noticing an almost one to one correspondence
between our process and $2 \gamma \rightarrow 2 \pi^{0}$. 
For the latter, the initial state photons produce intermediate 
quarks which decay into two pions via QCD interactions. Similarly,
the initial state gluons in the gluon fusion process generate 
techniquarks which decay into longitudinal Z-bosons. By the 
equivalence theorem \cite{EQ}, the longitudinal gauge bosons
may be treated as Nambu-Goldstone bosons (NGB).

The process $ 2\gamma \rightarrow 2\pi^{0} $ was first studied 
by Bijnens and Cornet \cite{R2}, and
independently by Donoghue et.al. \cite{R3} to one-loop level in chiral
perturbation theory. 
 In this talk we use  one-loop chiral perturbation to 
estimate the cross section for
$ 2g \rightarrow 2Z_{L} $. We note that the cross section was calculated 
in ref. \cite{RR1} in chiral limit in which all the pseudo NGB masses are 
zero. We find in generic TC models the pseudo NGB mass effect is important,
and it can give more than 200 \% correction over the chiral limit result.

\section{Chiral Lagrangian}

Let us consider one-generation model. 
The model has a $SU(8)_{L} \times SU(8)_{R}$ chiral symmetry which 
breaks down to
 $SU(8)_{V}$, generating 63 pseudo-NGBs. As a result of the
 symmetry breaking, technifermions obtain approximately degenerate 
 masses. The pseudo-NGBs comprise the  color singlets
 \be
 \Pi^{a}, \tilde{\Pi}^{a}, \Pi^{D},
 \ee
 and the 36  colored ones
 \be
 \Pi^{\alpha}, \Pi^{a \alpha},\Pi^{\mu i},\overline{\Pi}^{\mu i},
 \ee
 where $a$ and $\alpha$ are the SU(2) isospin and color octet indices
 respectively.  The $\Pi^{a}$s are the NGBs eaten by the weak
 gauge bosons.

The chiral Lagrangian in our model coupled with gluons is given
in the form
\be
{\cal L} = \frac{ F_{\pi}^{2}}{4} Tr \left( D_{\mu} 
U^{+} D_{\mu}U \right)
+ {\cal L}_{m} + {\cal L}_{4} + \cdots,
\label{r4}
\ee
with
\be
{\cal L}_{m} = \frac{1}{2} Tr \left( M ( U + U^{+} ) \right),
\ee
where $M$ is the mass matrix.
The form of ${\cal L}_{4}$ is a little complicated,
comprising terms involving four derivatives. The exact form may 
be found in  standard reference \cite{R6}.
Here $F_{\pi}$ is the technipion decay 
constant which determines the electroweak symmetry breaking
scale and is given by
\be
{}F_{\pi} = \frac{v}{2} \approx 123\, GeV,
\ee
and
\bear
&& U = \exp\left(i \frac{ \Pi \cdot T}{ F_{\pi} }\right), \nonumber \\
&& D_{\mu}U = \partial_{\mu} U + [G_{\mu},U], \hspace{.25in}
G_{\mu}= \frac{i g_{s}}{\sqrt{2}} G_{\mu}^{\alpha} T^{\alpha},
\label{r7}
\eear
where $T^{A}$, $A=1, \cdots, 63$, are the $SU(8)$ generators normalized
to
\be
Tr \left( T^{A} T^{B} \right) = 2 \,\delta^{AB},
\ee
and $g_{s}, G_{\mu}^{\alpha}$ are the strong coupling constant and
the gluon fields respectively.

Substituting (\ref{r7}) into (\ref{r4}) and 
expanding it, we have the desired
Lagrangian
\be
{\cal L} = {\cal L}_{0} + {\cal L}_{G-\Pi} + 
{\cal L}_{m}' + {\cal L}_{\Pi},
\ee
where $ {\cal L}_{0} $ is the free Lagrangian of
 the pseudo-NGBs.
 The $ {\cal L} _{G-\Pi}$ describes
pseudo-NGB coupling to gluons, and $ {\cal L}_{m}'$, ${\cal L}_{\Pi}$  
comprise four point interaction terms of pseudo-NGBs  
from the mass term  and the first term in (\ref{r4}) respectively.
They are given, in the unit $F_{\pi}=1$, by
\bear
{\cal L}_{G-\Pi} = && g_{s} G_{\sigma}^{\alpha} \left\{ 
- f^{\alpha\beta\gamma}
 \left( \Pi^{\beta} \partial_{\sigma} \Pi^{\gamma}
+ \Pi^{a \beta} \partial_{\sigma} \Pi^{a \gamma} \right) 
 - \frac{i}{2} \lambda_{ij}^{\alpha} \left( 
\Pi_{+}^{\mu i} \partial_{\sigma} \Pi_{-}^{\mu j} 
-\Pi_{-}^{\mu j} \partial_{\sigma} \Pi_{+}^{\mu i} \right) 
\right.\nonumber \\
&& \left.+ \frac{1}{24} \left( 2 
f^{\alpha\beta\gamma} z^{2} \delta_{T}^{ab} 
\Pi^{a \beta} \partial_{\sigma}
\Pi^{b \gamma} +
 i \lambda_{ij}^{\alpha} 
\delta_{T}^{\mu\nu} z^{2} 
\left( \Pi_{+}^{\mu i} \partial_{\sigma} \Pi_{-}^{\nu j} 
- \Pi_{-}^{\nu j} \partial_{\sigma} \Pi_{+}^{\mu i} \right)
 \right)\right\} \nonumber \\
&& + \frac{g_{s}^{2}}{8} G_{\sigma}^{\alpha}
 G_{\sigma}^{\beta} \left\{
4 f^{\alpha\gamma\tau} f^{\beta\delta\tau} 
\left( \Pi^{\gamma} \Pi^{\delta} + \Pi^{a \gamma} \Pi^{a \delta}
 \right) + \left\{ \lambda^{\alpha},\lambda^{\beta}\right\} 
_{ij} \Pi_{+}^{\mu i} \Pi_{-}^{\mu j} \right. \nonumber \\
&& \left. - \frac{z^{2}}{3} f^{\alpha\gamma\tau} 
f^{\beta\delta\tau} \delta_{T}^{ab}        
\Pi^{a \gamma} \Pi^{b \delta} -\frac{z^{2}}{12} 
\left\{ \lambda^{\alpha},\lambda^{\beta}\right\}_{ij}  
\delta_{T}^{\mu\nu} \Pi_{+}^{\mu i} \Pi_{-}^{\nu j}  \right\}.
\label{r11} 
\eear
\bear
{\cal L}_{\Pi} &=& \frac{1}{24} \left\{ 
   \partial_{\sigma} z^{2} \left( \delta_{T}^{ab} 
  \partial_{\sigma} \Pi^{a \alpha} \Pi^{b \alpha} 
   + \delta_{T}^{\mu\nu} \left(
   \partial_{\sigma} \Pi^{\mu i} \Pi^{\nu i} + 
   \partial_{\sigma} \overline{\Pi}^{\mu i} \overline{\Pi}^{\nu i} 
   \right) \right) \right. \nonumber \\
   && -\left(  \partial_{\sigma} z \right)^{2}  \left( \delta_{T}^{ab} 
   \Pi^{a \alpha} \Pi^{b \alpha} + \delta_{T}^{\mu\nu} \left(
   \Pi^{\mu i} \Pi^{\nu i} + 
    \overline{\Pi}^{\mu i} \overline{\Pi}^{\nu i} \right) \right) 
   \nonumber \\
   &&\left. -z ^{2}  \left( \delta_{T}^{ab} 
   \partial_{\sigma} \Pi^{a \alpha}  \partial_{\sigma} \Pi^{b \alpha} 
   + \delta_{T}^{\mu\nu} \left(
   \partial_{\sigma} \Pi^{\mu i}  \partial_{\sigma}\Pi^{\nu i} + 
   \partial_{\sigma} \overline{\Pi}^{\mu i}  
   \partial_{\sigma}\overline{\Pi}^{\nu i} \right) \right) \right\}.
   \label{r12}
\eear
\bear
{\cal L}_{m}' 
&=& \frac{z^{2}}{96} \left( 3 
Tr \left(  M  (\Pi^{\alpha})^{2}  \right)  + 
(2 \delta^{ab} + \chi^{ab})
Tr \left(  M  \Pi^{a \alpha} \Pi^{b \alpha} \right) 
+ \right. \nonumber \\  
&& \hspace{.25in} \left. (2 \delta^{\mu\nu} + \chi^{\mu\nu})
Tr \left(  M \left(  \Pi^{\mu i} \Pi^{\nu i} +\overline{\Pi}^{\mu i} 
\overline{\Pi}^{\nu i}  \right) \right) \right)  \nonumber \\ 
 &=&   
\frac{z^{2}}{48} \left( 3 m_{\alpha}^{2} (\Pi^{\alpha})^{2}   + 
 m_{a \alpha}^{2} (2 \delta^{ab} + \chi^{ab})
 \Pi^{a \alpha} \Pi^{b \alpha}  + \right. \nonumber \\  
&& \hspace{.25in} \left. m_{\mu i}^{2} (2 \delta^{\mu\nu} 
+ \chi^{\mu\nu})
\left(  \Pi^{\mu i} \Pi^{\nu i} +\overline{\Pi}^{\mu i} 
\overline{\Pi}^{\nu i}  \right) \right),
\label{r13}
\eear
where 
\be
\Pi_{\pm}^{\mu i} = \frac{1}{\sqrt{2}} 
\left(\Pi^{\mu i} \mp i \overline{\Pi}^{\mu i}\right),
\ee
 and $\Pi^{A}$ inside the trace should 
be regarded as $ \Pi^{A} \cdot T^{A}$.
 In deriving (\ref{r13}), we have used
the fact that $(T^{a})^{2} \sim I$, and thus commute with all the
generators, and 
\be
Tr \left( M (\Pi^{A})^{2} \right) = 2 m_{A}^{2}\, (\Pi^{A})^{2},       
\ee
which is from the quadratic mass terms of pseudo-NGBs. Note that
$ {\cal L}_{m}'$ is completely parameterized 
in terms of the pseudo-NGB
masses, independently of the detailed form of $M$.  
We also note that in (\ref{r11}), (\ref{r12})
 and (\ref{r13}),  only terms
 relevant for our process are presented
and $z \equiv \Pi^{3}$ is the NGB eaten by Z-boson.
The tensors are
defined by
\bear
&&\delta_{T}^{ab} = \mbox{Diag}( 1,1,0), \delta_{T}^{\mu\nu}
 = \mbox{Diag}(0,1,1,0)
\nonumber \\
&& \chi^{ab} = \mbox{Diag}( -1,-1,1), \chi^{\mu\nu} = 
\mbox{Diag}(1,-1,-1,1).
\eear

\section{$Z_{L}$-pair production rate}

With the Lagrangian given in sec.2, it is straightforward to 
calculate  the one-loop amplitude for 
$ 2g \rightarrow 2Z_{L} $.
Since the details of calculation  are very similar to those for
$ 2\gamma \rightarrow 2\pi^{0}$ and they can 
be found in ref.\cite{R3}, we present here only the
final result for 
$ G_{\mu}^{\alpha}(q_{1})G_{\mu}^{\alpha}(q_{1}) \rightarrow 2 z $,
which is given by
\be
A_{\mu\nu}^{\alpha\beta}(q_{1},q_{2}) = 
 \frac{g_{s}^{2}}{F_{\pi}^{2}}
\left( \frac{-i}{16 \pi^{2}} \right) 
\delta^{\alpha\beta} \left(
\frac{ g_{\mu\nu} q_{1}\cdot q_{2} - 
q_{2\mu}q_{1\nu} }{q_{1}\cdot q_{2}}\right)
\cdot {\cal A}(s), 
\label{r17}
\ee
where 
\bear
{\cal A}(s) &=& \frac{3}{4} m_{\alpha}^{2}
 \left( 1+ 2 I( m_{\alpha}, s)\right) 
\nonumber \\ 
& & + \frac{3}{2} \left( s + \frac{1}{6} 
m_{a \alpha}^{2} - \frac{2}{3} m_{z}^{2} \right)
 \left( 1+ 2 I( m_{a \alpha}, s)\right)  \nonumber \\
& & + \frac{1}{2} \left( s + \frac{2}{3} 
( m_{\mu i}^{2} -  m_{z}^{2}) \right)
 \left( 1+ 2 I( m_{\mu i}, s)\right),  
\label{r18}
\eear
with  $s = ( q_{1} + q_{2})^{2}$ and
\bear
I( m, s) &=& \int_{0}^{1} \frac{m^{2}}{x y s - m^2 + i \epsilon} \theta
(1-x-y) d x d y \nonumber \\
&=& \frac{m^{2}}{2 s} \left( \ln \left( 
\frac{ 1 + \sqrt{ 1-\frac{4 m^{2}}{s}}}{
1 - \sqrt{ 1-\frac{4 m^{2}}{s}}}\right) - i \pi \right)^{2} 
\hspace{.1in}
\mbox{for} \hspace{.1in} s > 4 m^{2},  \nonumber \\
&=& -\frac{m^{2}}{2 s} \left( \pi - 2 \arctan 
\sqrt{ \frac{4 m^{2}}{s} -1 } \right)^{2} \hspace{.1in}
\mbox{for} \hspace{.1in} s < 4 m^{2}.
\eear

\begin{figure}
\vspace{3.5in}
\caption{The cross sections for $Z_{L}$-pair and $W_{L}$-pair production
via gluon fusion in proton collisions at $E_{c.m.}= 14 TeV$. The solid lines  are for
the $q \bar{q}$ backgrounds, and dotted, dot-dashed, and dashed
lines are for the pseudo Nambu-Goldstone boson mass
 of $250 \,GeV$, $300 \, GeV$, and $350 \, GeV$ respectively. The thick
\noindent dot-dashed lines are for the chiral limit.
}
\end{figure}

The cross section using the fitted parton 
distribution functions for MRSA \cite{MRSA}
is plotted in Fig.1 in the range $200\, GeV \leq
 \sqrt{\hat{s}} \leq 1 \, TeV$ with the rapidity
 cut $|y| \leq 2.5$, and 
 over the standard model 
background for several values of the pseudo-NGB masses.  For
simplicity, 
we put identical values for the colored pseudo-NGB masses. 
The main sources of the background are  the quark fusion $
q \bar{q} \rightarrow ZZ$ and the gluon fusion $ gg \rightarrow ZZ$ via 
fermionic one-loop interactions. The quark fusion is the dominant
background at LHC energies as shown in ref.\cite{R1}, and it is about
 three times stronger than the gluon fusion in the invariant mass range
considered above.  In Fig.1a we
observe that our signal at energies below the pseudo-NGB threshold
is negligible relative to the background, but above the threshold 
larger than the background by a factor of $O(7-40)$ depending on the
energy. Notice that the pseudo NGB mass effect is still
important at energies above the pseudo NGB threshold
 and it can enhance the cross section by a facor two. 
 As expected, the signal   to background ratio becomes larger as the
invariant mass increases. This is mainly because
  the four-point vertices
of pseudo NGBs from the lagrangian ${\cal L}_{\Pi}$
in (\ref{r12}) are proportional to the invariant mass.
 Since our signal is so strong, it will increase the overall  
Z-boson pair production rate above the pseudo-NGB threshold, 
and this obviates the need to measure the polarization 
of the final state Z-bosons.

We also plot the cross section for $W_{L}^{+}W_{L}^{-}$ pair
production via gluon fusion. Because of the $SU(2)$ isospin
symmetry the amplitude for $gg \rightarrow W_{L}^{+}W_{L}^{-}$
is exactly given by Eq. (\ref{r18}) with $m_{z}$ replaced by
$m_{w}$, and  the cross section  is approximately
twice that of $Z_{L}$
pair production.  The dominant background for the $W_{L}$
pair production is also the $q \bar{q}$ fusion. From Fig.1b
we see that the signal above the colored pseudo NGB threshold
is stronger than the background by a factor of $O(6-40)$.
Notice that we have applied a more stringent rapidity cut, $
|y| \leq 1.5 $, to enhance the signal to background ratio.

Clearly the signal should be observable without difficulty
at LHC with the planned
c.m. energy $\sqrt{s} =14\, TeV$ and the integrated luminosity
$100 fb^{-1}$ per year. As an example, let us consider the 
$Z_{L}$-pair production via gluon fusion with the pseudo NGB mass of
$300 \, GeV$. The integrated cross section with the invariant
msss above the pseudo NGB threshold is $\sigma(\sqrt{\hat{s}}
\geq 600 \,GeV) = 3.4\,pb $. The most clean signal for the $Z_{L}$-pair
production would be four leptons of electrons and muons without
jets. With the branching ratio of $.45 \%$, the event rate 
would be $1530$ per year.  Instead if the signal is two leptons
 of electrons or
muons with missing mass, then the event rate would be $9180 $
per year.
This shows that our process could be an effective
probe of  SBS  of TC type with pseudo NGBs having 
nonzero color and electroweak isospin.

\vspace{.18in}
\noindent  I am greatly indebted to Estia Eichten for many illuminating
suggestions and comments.

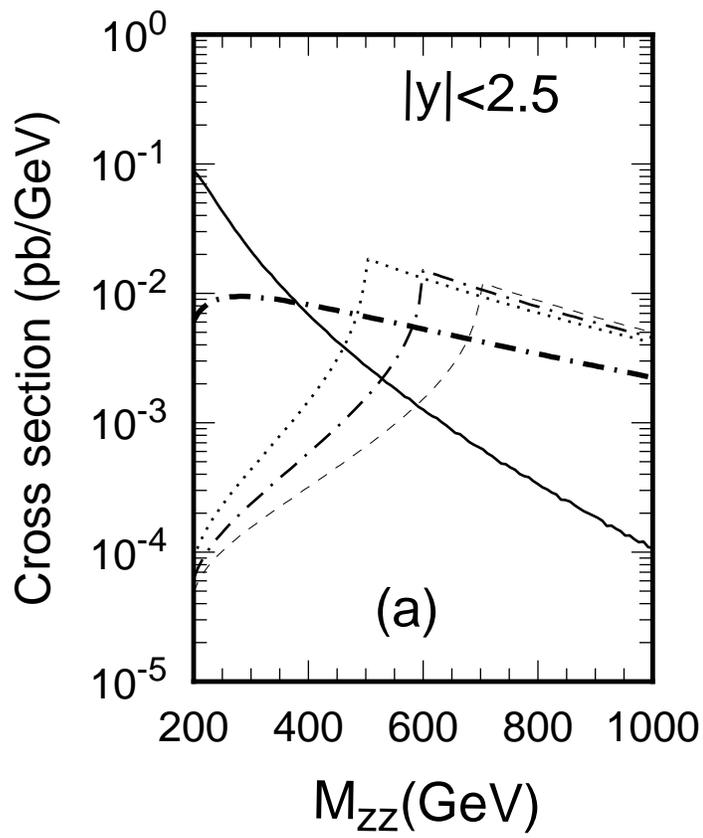
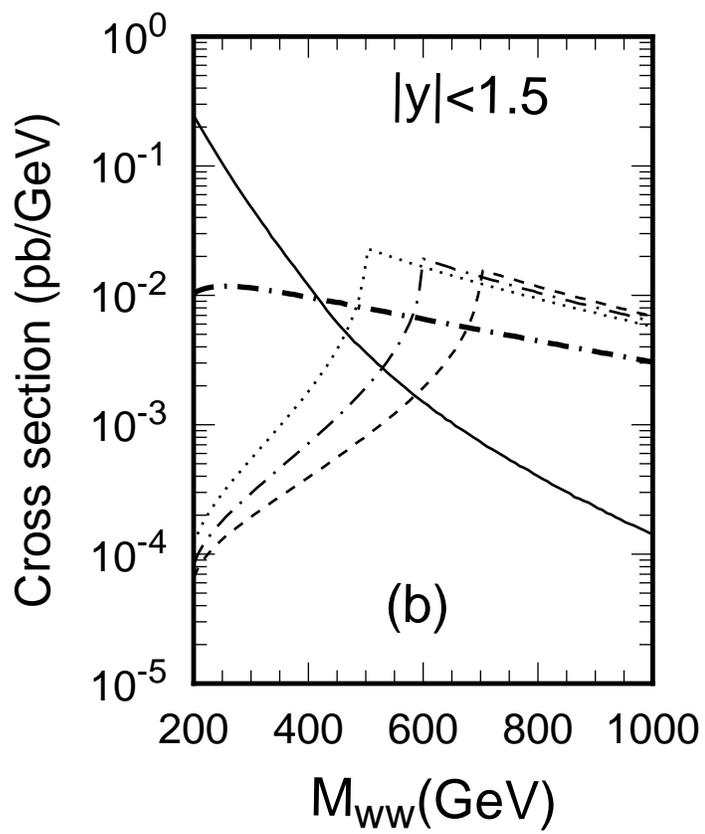
\end{document}